\documentclass[12pt]{article}

\newcommand{\bs}{\begin{singlespace}}
\newcommand{\es}{\end{singlespace}}

\usepackage{doublespace}
\usepackage{graphicx}

\begin{document}

\bs

\begin{center}\large
Polarization Correlations of $^1S_0$ Proton Pairs as Tests of 
Bell and Wigner Inequalities
\vspace{.5cm}

\normalsize C. Polachic, C. Rangacharyulu \\ \small Department of Physics and 
Engineering Physics \\ University of Saskatchewan, Saskatoon, Saskatchewan, Canada 
S7N 5E2 \\ \vspace{.5cm} \normalsize A. van den Berg, M. Harakeh, M. de Huu, H. 
W\"{o}rtche \\ \small Kernfysisch Versneller Instituut, \\ Zernikelaan 25, NL-9747 
AA Groningen, The Netherlands 
\\ \vspace{.5cm} \normalsize J. Heyse \\ \small Vakgroep Subatomaire en 
Stralingsfysica, \\ Rijksuniversiteit Gent, Proeftuinstraat 86, B-9000 Gent, 
Belgium \\ \vspace{.5cm} \normalsize C. B\"{a}umer, D. 
Frekers \\ \small Institut f\"{u}r Kernphysik, Westf\"{a}lische 
Wilhelms-Universit\"{a}t, \\ Wilhelm-Klemm-Str. 9, D-48149 M\"{u}nster, Germany \\ 
\vspace{.5cm} \normalsize J. 
Brooke \\ \small Department of Mathematics and Statistics \\ University of 
Saskatchewan, Saskatoon, Saskatchewan, Canada S7N 5E6

\begin{abstract} In an experiment designed  to overcome the loopholes of observer
dependent reality and satisfying the counterfactuality condition, we
measured polarization correlations of $^1S_0$ proton pairs produced in
$^{12}C(d,^2He)$ and $^ 1H(d,^2He)$ reactions in one setting. The results of
these measurements are used to test the Bell and Wigner inequalties
against the predictions of quantum mechanics.\end{abstract}

\end{center}

\normalsize
\section{Introduction}
Nearly seven decades have passed since  the appearance of the seminal
publication by Einstein, Podolsky
and Rosen \cite{epr} and the question of completeness of physical
reality in quantum mechanics is yet  unsettled.  It is well known that
Bell \cite{bell} suggested a quantitative means to test the
predictions of local hidden variable theories against those of quantum
mechanics.  He considered  an entangled pair of spin 1/2 particles in the
$^1S_0$ state.  He derived  results for spin  correlations of this
pair along four arbitrary axes in space from the quantum mechanical 
reasoning and that of local hidden variable theories.  The main result
was that for a judicious choice of axes, the quantum mechanical
predictions are incompatible with Bell's inequality and thus experiments 
would  be able to decide between the two.  It should be remarked that 
with a couple of exceptions, the experiments in the ensuing period
were carried out with pairs of photons (spin 1 objects), rather than 
spin 1/2 pairs.  See Vaziri et al \cite{zeil} for the latest results.

A less well-known but, perhaps, equally important result was obtained
by Wigner \cite{wig}.  He evaluated  the spin correlations of a  pair of
particles along three axes, instead of four, from the 
hidden variable estimates and those of quantum mechanics. To our
knowledge, these predictions have never been tested.

In this letter, we present first results of our measurements on pairs
of protons.  In the literature, there is only one such experiment
which employed spin correlations due to elastic proton scattering off
a proton target \cite{lrm}.  This pioneering experiment nevertheless
had some drawbacks which the authors, themselves, recognized.  Our
motivation was to minimize or avoid those loopholes while providing a
first-ever test of the Wigner inequality.  

\section{Experimental Setup}

The measurements were carried out using the 170 MeV  deuteron beams from the
AGOR cyclotron facility of  KVI, Groningen, the Netherlands. The experimental
  setup is well described in ref. \cite{kvi}. \begin{figure}[htb]\begin{center} 
\includegraphics[width=.6\textwidth] 
{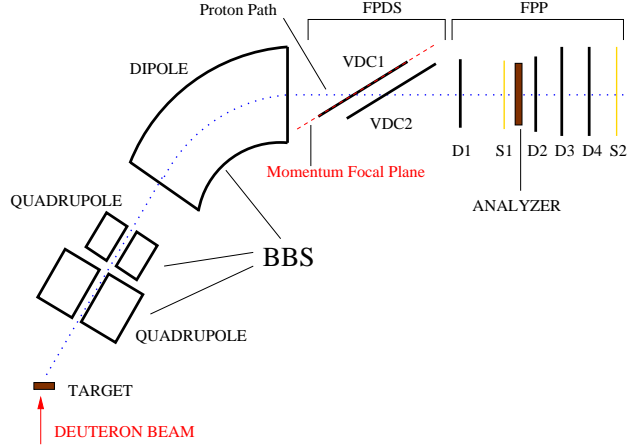}\end{center} 
\caption{Schematic diagram of the experimental arrangement (top 
view).}\label{apparatus}\end{figure} Carbon targets
are employed to prepare entangled $^1S_0$  two proton states  by
$^{12}C (d,^2He) ^{12}B^*$ and $^1H(d,^2He) n$ reactions, the latter
arising 
from the hydrogen impurity in the target. The layout
of the detector system is shown in Figure \ref{apparatus}.  From 
several reactions that occur, the Big Bite Spectrometer (BBS) selects
positively charged particles of momenta $p = 600 \pm 50$ MeV/c.
The spectrometer is equipped with a focal plane detector system (FPD)  which
consists of two planes of drift chambers (VDCs) to determine the momentum of the 
protons.  Behind this 
system is the focal plane polarimeter (FPP), which consists of four
sets of multi-wire proportional chambers (MWPCs), labeled D1-D4, and two sets of 
plastic scintillator
arrays (S1 and S2).  A carbon analyzer is placed just next to
scintillator S1. 

A particle of positive charge and momentum 600 $\pm$ 50 MeV/c
entering the spectrometer will leave ionization trails in the
VDCs.  This causes scintillations in S1.  The particle undergoes spin-dependent 
scattering
in the carbon analyzer, leaves ionization trails in the wire chambers
and passes through S2.  For an event to be registered, we require that
a particle would have passed all the way up to S2.  The scintillators
have an intrinsic time resolution better than 1 ns.

The four momentum vector of the particle is determined by measuring the position 
and 
direction of its trajectory in the BBS focal plane by means of the high resolution 
VDCs and chamber D1.  The particle trajectory downstream the analyzer is measured 
by 
means of chambers D2-D4.  The redundancy of three sets of spatial coordinates 
upstream and downstream the analyzer ensures a flat detection efficiency.

The data taking logic is set as follows:  A hardware coincidence 
requires that there is a signal between a scintillator in S1 and
another in S2, within a time interval of 150 ns.  The window is set
wide open on purpose.  First, the protons travelling along various  paths
in the spectrometer have different flight times between S1 and S2.
More importantly, the cyclotron operates at a frequency of 43 MHz.
Taking data in this mode provides a good handle on random 
coincidences during the off-line analysis, since events separated by
more than 20 ns arise due to random coincidences.  If this coincidence
condition is satisfied, the data from all wire chambers are read out and subject to 
an online data analysis performed by a set of fast digital signal processors.  
Events 
fulfilling the requirements for two coincident protons passing the setup are stored 
on tape.  The TDC data is stored in bins of 1 ns and allow for an event 
registration 
over 350 ns, to permit
the ion drift times to be recorded.  Protons, travelling at speeds of 
about 0.5c, spend a few nanoseconds in the FPD region and another 
few nanoseconds in the FPP region.  The transit times of the
ionization
trails in the VDCs and wire chambers are of the order of few hundred
nano
seconds (drift speeds are of 50 $\mu$m/ns, and drift lengths are
about 5 mm).  The protons have been out of the entire system long
before the trail information
is received by the electronic circuitry and processed by the data 
acquisition system.  Since a proton spends a little over
10 picoseconds at each detection element, which generate their
signals independent of each other, we believe that the communication
between pairs of protons by means other than superluminal signals is
not possible.

This standard nuclear physics 
instrumentation allows the data  of each particle track to be 
recorded providing information on particle identification, 
its momentum vector, its time relation with 
respect to other particles in the same event, and its direction after it
passes through the carbon analyzer.  With the momentum axis of the
protons as the reference z-axis, the polarization analysis can be done
for any arbitrary set of  x-y axes in the analyzer plane, with an 
acceptance of the entire 180 degrees.  We eliminate what 
is known as trigger bias in subatomic physics and address two
loopholes raised by Peres \cite{peres} and Redhead\cite{red}:
``counterfactuality'' and ``conscious-observer dependence" of measurements.  
As we cover the entire angular range for polarization analysis in the x-y plane, 
the 
limitation of 
a single angular setting per particle per event does not arise here.
Also, since the orientation of axes is done during the off-line
analysis several months after the experiments have been performed,
we do not have the problem of the ``choice of analyser setting made
well in advance of the emission of the particles from the source''\cite{red}.

\section{Measurement and Data Analysis}

\begin{figure}[htb]\begin{center} 
\includegraphics[width=.6\textwidth] 
{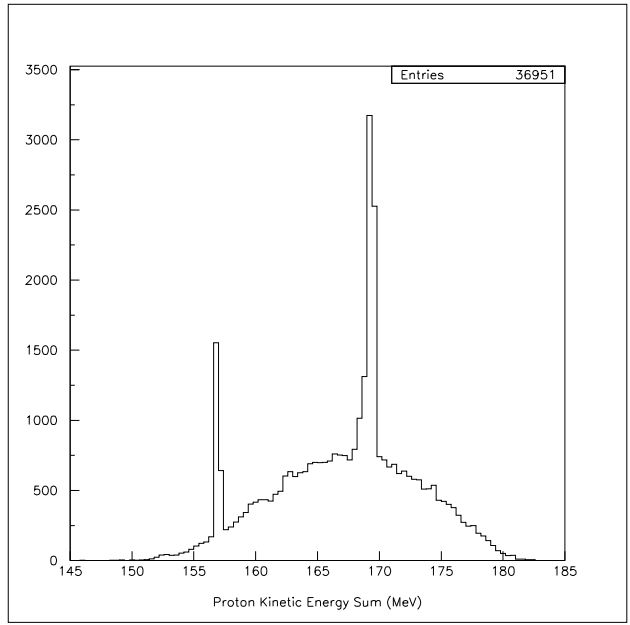}\end{center} 
\caption{Sum of the proton-pair kinetic energies in the $(d,^2He)$ reaction on 
$^{12}C$ and $^1H$ for $E_d = 170$ MeV.  The singlet peaks above 
background are clearly seen.}\label{spectrum}\end{figure}

In the summer of 2001, we ran the experiment over a three-day period. 
From a data volume of several tens of millions of triggers, we could
identify 4.7 million as two-proton events.  The next task
was to select  the events due to the $(d,^2He)$ reaction.  We
know that, in this reaction,  protons in each  pair are at rest
relative to each other in the center of mass frame.  We thus
apply standard relativisitic kinematical equations with each event
Lorentz boosted to the center of mass frame, and select the events
for which the relative kinetic energy of the proton pair is less than
1 MeV, allowing for the instrumental resolution.   Figure \ref{spectrum} 
 is a
plot of the sum of the kinetic energies of two protons for events thus selected.  
Noteworthy are the peaks 
at 170  and 158 MeV, respectively due to $^1H(d,^2He)$ and $^{12}C(d,^2He)^{12}B$ 
reactions.  The events
in these sharp peaks are thus proton pairs  due to 
the decay of a $^2He$ $^1S_0$ intermediate state.

We follow the fate of each proton in each event as they pass through
the carbon analyzer, acquiring the information in the detector systems
D1-D4 and S1, S2.  First of all, we  use only those  events where 
both protons  scatter at angles larger than 3 degrees  in the carbon
analyzer, since the most forward scattering is Coulomb type and it  is not
spin-dependent.  We have a good estimate of the carbon analyzing power
from earlier works of KVI groups \cite{anal}.  It was estimated to be
0.2 for angles 5-20 degrees.  We realized that our data provides an 
inherent measurement of the analyzing power since we deal with proton
pairs in which  spins of the two protons in each 
pair are oppositely oriented.   Using this information, we were able
to deduce the analyzing power for our dynamical range as  $A=0.25$.
It is quite satisfying to note that this result is in good agreement with
the earlier KVI results. 

\section{Results}

We are now in a position to deduce the experimental correlations to
compare against the expectations from conventional quantum mechanical
arguments and those of Bell\cite{bell} and Wigner\cite{wig}. 

\subsection{Bell correlations}

The formulation of Bell's inequality finds its basis in the EPRB {\em gedanken} 
experiment of Bohm\cite{bohm}, considering a pair of spin-half 
particles in the singlet state.

A measurement is made on each particle, determining the projection of its intrinsic 
spin vector along some arbitrary direction orthogonal to the momenta of the pair.  
Bell showed that the products of such measurements lead to an 
experimental condition under which the existence of the hidden parameters could be 
tested against the completeness of quantum theory.

If $\hat{a}$ and $\hat{a^{\prime}}$ are arbitraty unit vectors along which the spin 
of the first particle is measured (and $\hat{b}$ and $\hat{b^{\prime}}$ apply 
similarly to the second particle), the hidden variables framework results in an 
algebraic condition for the probabilities, $P(\cdot,\cdot)$:
\begin{equation} |P(\hat{a},\hat{b}) - P(\hat{a},\hat{b^{\prime}}) + 
P(\hat{a^{\prime}},\hat{b}) +
P(\hat{a^{\prime}},\hat{b^{\prime}})| \leq 2.\label{delta} \end{equation}

\noindent while the quantum expectation value, $P_{QM}(\theta)$, for these 
correlation 
functions is 
\begin{equation} 
P_{QM}(\theta) = -\cos\theta \label{qmexpect} \end{equation} where $\theta$ is the 
angle between the unit vectors $\hat{a}$ and $\hat{b}$.  (\ref{delta}) is satisfied 
for the hidden variables case, but (for certain choices of the co-planar unit 
vectors) it is violated by (\ref{qmexpect}).

\begin{figure}[htb]\begin{center} 
\includegraphics[width=.5\textwidth] 
{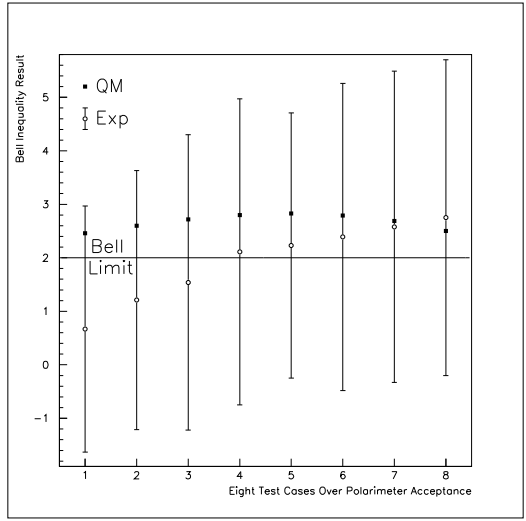}\end{center} 
\caption{Eight test cases of the experimental data from Table \ref{belltable} 
versus quantum predictions and the limit of the Bell 
inequality.}\label{bellfig}\end{figure}

During the off-line analysis (one year after the experiment was performed), we 
selected eight angular combinations for which the quantum
mechanical predictions exceed the Bell limits.  The results of this analysis are 
shown 
in table \ref{belltable} and figure \ref{bellfig}.

The first observation is the large error due to limited statistics associated with 
the data.  Secondly, the quantum mechanical results for all sets are nearly
constant, while the data seems to vary from 0.7 (consistent with Bell's limit) to 
about 2.75 (clearly violating the inequality).  Needless to
say, definitive conclusions await more precise measurements.  However, we would 
like 
to emphasize that this type of measurement has an
excellent potential to settle this problem.

\begin{table}\begin{center}\begin{tabular} {|c|c|c|r@{$\pm$}l|}
\hline
 & Bell & Quantum & \multicolumn{2}{|c|}{} \\
Case & Inequality & Mechanics & \multicolumn{2}{|c|}{Experiment} \\ \hline
1 & $P(0^{\circ},25^{\circ}) - P(0^{\circ},75^{\circ}) + P(50^{\circ},25^{\circ}) + 
P(50^{\circ},75^{\circ})$ & 2.46 & 0.67 & 2.30 \\
2 & $P(0^{\circ},30^{\circ}) - P(0^{\circ},90^{\circ}) + P(60^{\circ},30^{\circ}) + 
P(60^{\circ},90^{\circ})$ & 2.60 & 1.21 & 2.42 \\
3 & $P(0^{\circ},35^{\circ}) - P(0^{\circ},105^{\circ}) + P(70^{\circ},35^{\circ}) 
+ 
P(70^{\circ},105^{\circ})$ & 2.72 & 1.54 & 2.76 \\
4 & $P(0^{\circ},40^{\circ}) - P(0^{\circ},120^{\circ}) + P(80^{\circ},40^{\circ}) 
+ 
P(80^{\circ},120^{\circ})$ & 2.80 & 2.11 & 2.86 \\
5 & $P(0^{\circ},45^{\circ}) - P(0^{\circ},135^{\circ}) + P(90^{\circ},45^{\circ}) 
+ 
P(90^{\circ},135^{\circ})$ & 2.83 & 2.23 & 2.48 \\
6 & $P(0^{\circ},50^{\circ}) - P(0^{\circ},150^{\circ}) + P(100^{\circ},50^{\circ}) 
+ 
P(100^{\circ},150^{\circ})$ & 2.79 & 2.39 & 2.87 \\
7 & $P(0^{\circ},55^{\circ}) - P(0^{\circ},165^{\circ}) + P(110^{\circ},55^{\circ}) 
+ 
P(110^{\circ},165^{\circ})$ & 2.69 & 2.58 & 2.91 \\
8 & $P(0^{\circ},60^{\circ}) - P(0^{\circ},180^{\circ}) + P(120^{\circ},60^{\circ}) 
+ 
P(120^{\circ},180^{\circ})$ & 2.50 & 2.75 & 2.95 \\ 
\hline \end{tabular}\end{center}\caption{Quantum predictions and experimental 
results for several violating cases of the Bell 
inequality.}\label{belltable}\end{table}

\subsection{Wigner Inequality}

The Wigner inequality is similar to Bell's, but arguably stronger, offering a 
clearer method of distinguishing between quantum predictions and hidden variables 
results.  This inequality is:
\begin{equation} P(a,c)_{corr} - P(a,b)_{corr} - P(b,c)_{corr} \leq 0 
\label{wigineq} .\end{equation}
$P(\cdot,\cdot)_{corr}$ is the probability that the spin projections of particles 
one and two are correlated (pointing in the same direction) along their respective 
measurement axes.  The three axes $\hat{a},\hat{b}$ and $\hat{c}$ are co-planar 
with $\hat{b}$ bisecting the other two.  Agreement of the experiment with 
(\ref{wigineq}), if observed, would be evidence 
for the existence of hidden variables, while a positive experimental value would be 
consistent with the predictions of quantum theory.

\begin{figure}[htb]\begin{center} 
\includegraphics[width=.5\textwidth] 
{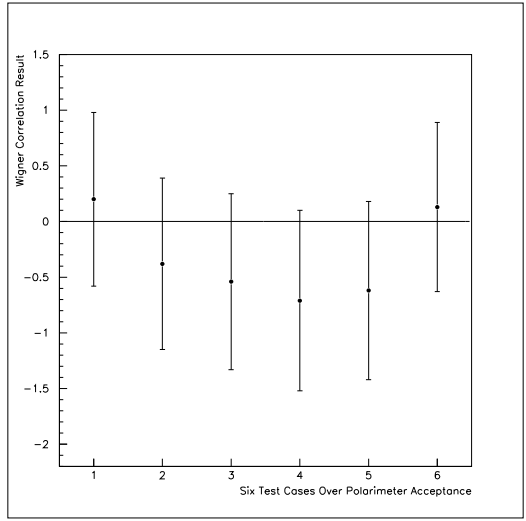}\end{center} 
\caption{Results of Wigner's inequality for the six cases of 
Table~\ref{wigtable}.}\label{wigfig}\end{figure}

We selected six sets of orientations, spanning the entire angular range $0^{\circ}$ 
to $180^{\circ}$.  The results are shown in table
\ref{wigtable}, and are shown in figure \ref{wigfig}.  As can be seen, while the 
data 
may be consistent with the quantum predictions due to
the large errors, there is a clear trend for the results to lie below the zero 
line, 
favoring the hidden variables condition.  As before, a
larger data sample is required to settle this question.

We suggest that for the first time, in a single setting we have obtained data to 
test both the Bell and Wigner inequalities, covering the entire
angular range, thus avoiding the loopholes of observer dependence and 
counterfactuality.  We plan to continue this work with similar tests in
the near future.

It is a pleasure to thank P. Busch for valuable discussions.  We acknowledge the 
financial support of KVI and NSERC Canada.

\begin{table}\begin{center}\begin{tabular} {|c|c|r@{$\pm$}l|}
\hline && \multicolumn{2}{|c|}{Experimental} \\ 
& Wigner Comparison & \multicolumn{2}{|c|}{Result} \\ \hline

1 & $P(0^{\circ},30^{\circ})-P(0^{\circ},15^{\circ})$ & 0.20 & 0.78 \\ & 
$-P(15^{\circ},30^{\circ})$ &  \multicolumn{2}{|c|}{} \\ \hline

2 & $P(0^{\circ},60^{\circ})-P(0^{\circ},30^{\circ})$ & -0.38 & 0.77 \\ & 
$-P(30^{\circ},60^{\circ})$ & \multicolumn{2}{|c|}{} \\ \hline

3 & $P(0^{\circ},90^{\circ})-P(0^{\circ},45^{\circ})$  
& -0.54 & 0.79 \\ & $-P(45^{\circ},90^{\circ})$ & \multicolumn{2}{|c|}{} \\ \hline

4 & $P(0^{\circ},120^{\circ})-P(0^{\circ},60^{\circ})$ & 
 -0.71 & 0.81 \\ & $-P(60^{\circ},120^{\circ})$ & 
\multicolumn{2}{|c|}{} \\ \hline

5 & $P(0^{\circ},150^{\circ})-P(0^{\circ},75^{\circ})$  & -0.62 & 0.80  \\ & 
$-P(75^{\circ},150^{\circ})$ & 
 \multicolumn{2}{|c|}{} \\ \hline

6 & $P(0^{\circ},180^{\circ})-P(0^{\circ},90^{\circ})$ & 0.13 & 0.76  \\ & 
$-P(90^{\circ},180^{\circ})$ & 
\multicolumn{2}{|c|}{} \\ \hline
\end{tabular}\end{center}\caption[Wigner Inequality Results]{Results for Wigner 
Inequality 
test.}\label{wigtable}\end{table}

\es

\begin{thebibliography}{00}

\bibitem{epr} A. Einstein, B. Podolsky and N. Rosen, Physical Review {\bf 47} 
(1935) 777.

\bibitem{bell}
J.S. Bell, Physics {\bf1} (1964) 195; Reviews of Modern Physics 
{\bf 38} (1966) 447.

\bibitem{zeil}
A. Vaziri, G.Weihs and A. Zeilinger, Physical Review Letters {\bf89}
(2002), 240401.

\bibitem{wig}
E. P. Wigner,  Am. J. Physics, {\bf38} (1970) 1005.

\bibitem{lrm}
M. Lamehi-Rachti and W. Mittig, Physical Review {\bf D 14} (1976) 2543.

\bibitem{kvi}
S. Rakers et al, Nuclear Instruments and Methods in Physics Research 
{\bf A481}  (2002) 253.

\bibitem{peres}
A. Peres, \emph{Quantum Theory:  Concepts and Methods}, Kluwer,
Dordrecht, (1995).

\bibitem{red}
M. Redhead, \emph{Incompleteness, Nonlocality, and Realism}, Clarendon,
Oxford (1987).

\bibitem {anal} V. Hannen, Doctoral Thesis, Westf\"{a}lische 
Wilhelms-Universit\"{a}t M\"{u}nster (2001).

\bibitem{bohm} D. Bohm, {\em Quantum Theory}, Prentice-Hall, Englewood Cliffs 
(1951).

\end{thebibliography}
\end{document}